\documentstyle[12pt]{article}

\catcode`\@=11
\begin{document}
\begin{titlepage}
\begin{centering}
\title{
Polarised parton distributions
\thanks{Partially supported by CONICET-Argentina.}}
\author{D.de Florian, \@\@ L.N.Epele, \@\@ H.Fanchiotti, \@\@ C.A.Garc\'{\i}a
Canal, \@\@
R.Sassot \\ \\ Laboratorio de F\'{\i}sica Te\'{o}rica \\ Departamento de
F\'{\i}sica \\
Universidad Nacional de La Plata \\ C.C. 67 - 1900 La Plata \\
Argentina}

\date{10 June 1994\\}
\maketitle
\end{centering}
\begin{abstract}
We analyze spin dependent parton distributions consistent with the most recent
measurements of the spin dependent deep inelastic scattering structure
functions
and obtained in the framework of the spin dilution model.
Predictions for the doubly polarised proton-proton Drell-Yan asymmetry,
for the high $p_{T}$ photon production mechanism and  $J/\Psi$ excitation are
calculated using these distributions and are shown to be particularly adequate
to unveil the polarisation of partons in the proton.
\end{abstract}
\end{titlepage}

\pagenumbering{arabic}
\noindent {\large \bf I. Introduction:}\\

Recently the SMC collaboration at CERN has reported on a measurement of the
proton polarised asymmetry in deep inelastic scattering of polarised muons off
polarised hydrogen \cite{SMC2}. This measurement was meant to corroborate and
extend the
controversial data produced by EMC in 1988 \cite{EMC} on the first moment of
the proton
spin dependent structure function $g_{1}^{p}$
\begin{equation}
\Gamma_{1}^{p}\mid_{EMC}=\int_{0}^{1}g_{1}^{p}dx=0.126 \pm 0.010(stat.) \pm
0.015 (sys.)
\end{equation}
which has been interpreted as an important contribution to the structure
function either from
the polarisation of gluons, of the strange quarks or of both.

Notwithstanding the new results for the asymmetry confirm those of the
previous measurement, the new values taken at small $x$ differ with the
extrapolations assumed in \cite{EMC} yielding
for the  moment of the spin dependent structure function
\begin{equation}
\Gamma_{1}^{p}\mid_{SMC}=0.136 \pm 0.011(stat.) \pm 0.011 (sys.)
\end{equation}
to be compared with Eq.(1). Part of the difference is also due to the fact
that $g_{1}^{p}$ is obtained from the measured asymmetry $A_{1}^{p}(x,Q^{2})$
through
\begin{equation}
xg_{1}^{p}(x,Q^{2})=\frac{A_{1}^{p}(x,Q^{2})F_{2}^{p}(x,Q^{2})}{2(1+R(x,Q^{2}))}
\end{equation}
and they have used a recent parametrization \cite{par}
for the spin independent structure function $F_{2}^{p}(x,Q^{2})$ which has a
better accuracy  at low $x$.

The rise in the value of the moment together with the fall in the theoretical
expectation for the quark contribution to the moment, resulting from the new
values for the
$F$ and $D$ parameters \cite{FyD}, reduces considerably the amount of gluon or
strange
quark polarisation needed to understand the data. This situation forces an
update of the parametrizations of quark and gluon spin dependent distributions
and also a reconsideration
of the experiments meant to size the gluon and strange quark polarisation.

In this paper we present two new sets for the quark and gluon spin dependent
distributions obtained including the recent SMC data on the proton in the
fitting procedure.
One with an important net gluon polarisation and another in
which this polarisation is negligible. Both are constructed in the framework of
the spin dilution model [6-7] and are in agreement with
all the avalilable deep inelastic scattering polarised asymmetries [2],[8-10].
The main features of the valence quark, sea quark, and gluon distributions in
both
sets are discussed. The low $x$ behaviour of the asymmetries and structure
functions resulting from the distributions is analyzed and special emphasis is
given to the comparison between  this behaviour and those assumed in the
extrapolation of the measured data. These extrapolations are key ingredients
in the estimation of the moments $\Gamma_{1}^{p}$, $\Gamma_{1}^{n}$, and
$\Gamma_{1}^{d}$.

Up to now, different experiments have been proposed in order to discriminate
between the alternative of a large net gluon polarisation in the proton and an
important contribution from the strange quarks \cite{Rey}. Predictions are made
using different
sets of distributions. In section III we calculate cross sections
for the doubly polarised proton-proton Drell-Yan asymmetry,
for the high $p_{T}$ photon production mechanism and  $J/\Psi$ excitation
using our updated sets and compare them with previous results.
In so doing, we find that not all the proposed mechanisms are actually able to
discriminate between the different scenarios.

We also comment on the correct implementation of the factorization scheme
in the use of the polarised distributions. This has not been taken into account
in some
proposals and is shown to ruin the alleged discriminative power of some
experiments.

Finally, we  show that certain experiments, suitably combined, can give
additional information such as the way in which the spin is distributed
among the different flavours in the sea. This information is clearly
beyond the scope of DIS measurements and has to be guessed in any
spin dependent parametrization, nevertheless is an important ingredient
in the knowledge of the proton structure.\\

\noindent {\large \bf II. Spin-dependent parton distributions:}\\

In the recent years, the increased precision and variety of data yielded by
unpolarised lepton-hadron and hadron-hadron experiments has considerably
improved our knowledge of parton distributions in the proton. Nowadays,
extremely precise unpolarised parton distributions are extracted from global
fits with a decreasing number of model assumptions about them. The situation,
however, is quite different for the spin dependent parton distributions for
which the data is comparatively scarce. The main experimental imput available
for the extraction of the spin dependent parton distributions, $\Delta
q_{i}(x,Q^{2})$ and
$\Delta G(x,Q^{2})$, are the spin dependent structure functions
$g_{1}^{N}(x,Q^{2})$  which are now known for the proton, the neutron and
the deuteron. The relation between parton distributions and structure functions
depends on the factorization scheme chosen to define the former. In what
follows we shall adopt the one in which the distributions are related to the
structure functions by \cite{AA}
\begin{equation}
g_{1}^{p}(x,Q^{2})=\frac{1}{2}\sum_{i}e_{i}^{2}\Delta q_{i}(x,Q^{2})-
\frac{<e_{i}^{2}>}{2}\frac{\alpha_{s}}{2\pi} n_{f}\Delta G(x,Q^{2})
\end{equation}
and not the DIS like scheme \cite{Rat}, labeled by a caret, where
\begin{equation}
g_{1}^{p}(x,Q^{2})=\frac{1}{2}\sum_{i}e_{i}^{2}\Delta \hat{q}_{i}(x,Q^{2})
\end{equation}
The relationship between them is simply given by
\begin{equation}
\Delta q(x,Q^{2})=\Delta \hat{q}(x,Q^{2})+\frac{\alpha_{s}}{4\pi}\Delta
G(x,Q^{2})
\end{equation}
where $\Delta q$ is related to the conserved part of the axial current.
The expression for $g_{1}(x,Q^{2})$ is known up to order $\alpha_{s}$.

As the present information is not enough to determine individualy the
distributions,
it is imperative to make some assumptions about them. For this we follow the
main
lines of the spin dilution model \cite{sdm}, implemented as in reference
\cite{sdpd},
which relates the spin dependent quark and gluon distributions with the
corresponding spin independent distributions by means of spin dilution
function.
This function is fixed in order to satisfy the constraints on the polarised
distributions for $x \rightarrow 0$ and $x \rightarrow 1$ leaving a few
parameters to be adjusted. Doing this, the full information about the
unpolarised parton distributions and the constraints on the polarised ones
are taken into account.

Let us obtain different sets of spin-dependent quark and gluon
distributions compatible with the deep inelastic scattering polarised
asymmetries available at the moment. To this end we follow the procedure
suggested in
ref. \cite{sdpd} which consists in fixing the free parameters of the spin
dilution model in the two following ways.

\noindent {\it i)} In this first set (LP1), the disagreement between the
Ellis-Jaffe
sum rule prediction \cite{EJ} for $\Gamma _{1}^{p}$
\begin{equation}
\Gamma _{1}^{p}\mid_{Ellis-Jaffe}=\frac{F+D}{12}[(1-\frac{\alpha_{s}}{\pi})
+\frac{1}{3}\frac{3F/D-1}{F/D+1}(5-\frac{\alpha_{s}}{\pi}(1-4C_{F}))]
=0.1766\pm 0.006
\end{equation}
and the experimental value Eq.(2) is ascribed, as it is commonly accepted
\cite{AA},
to the anomalous gluon contribution to $g_{1}^{p}$
\begin{equation}
\Delta \Gamma_{1}^{gluon}=\int_{0}^{1}dx
\frac{<e_{i}^{2}>}{2}\frac{\alpha_{s}}{2\pi}
n_{f}\Delta G(x)
\end{equation}
which means
\begin{equation}
\Gamma _{1}^{p}\mid_{Ellis-Jaffe}-\Delta \Gamma_{1}^{gluon}=\Gamma
_{1}^{p}\mid_{SMC}
\end{equation}
and implies
\begin{equation}
\Delta \Gamma_{1}^{gluon}=0.040
\end{equation}
As we have previously mentioned, this value is considerably lower than the one
obtained in ref. \cite{sdpd} because of the new values for $\Gamma_{1}^{p}$
found by SMC and
the most recent hyperon $\beta$-decay data \cite{FyD} parametrized in terms of
$F$ and $D$.
Once the spin dilution
parameter for the gluon is fixed in order to satisfy Eq.(7), the other
parameters
of the model are fixed just to reproduce the known asymmetries and the
quantities $F+D$ and $F/D$.

\noindent {\it ii)} Another extreme situation is given by the second set (LP2),
in which $\Delta \Gamma_{1}^{gluon}$
is forced to be zero. In this set, the discrepancy is ascribed to a negative
polarisation in the strange quark sea, namely
\begin{equation}
\Delta \Gamma_{1}^{s}=\frac{1}{3}(1-\frac{\alpha_{s}}{\pi}C_{F})\int_{0}^{1}dx
\Delta s
\end{equation}
fixed by
\begin{equation}
\Gamma _{1}^{p}\mid_{Ellis-Jaffe}+\Delta \Gamma_{1}^{s}=\Gamma
_{1}^{p}\mid_{SMC}
\end{equation}
This equation forces the strange quark contribution to be negative at variance
with those of the non strange sea quarks, which in the framework of the spin
dilution model are assumed to be radiatively generated from the valence quarks
and thus positive. The implications of this will be adressed in the next
section.

Figures (1-3) show the asymmetries for proton, neutron and deuteron as
calculated
with set LP1 against the measured values. The second set yields similar results
for the asymmetries. Table (1) shows the spin dilution model parameters for
both sets coming from a global fit to all the available data
using the set MRSD$_{-}$  given in reference
\cite{MRS}, which agree with the most recent HERA results,
for the unpolarised parton distributions. The parameters are fixed for
$Q^{2}=10\,GeV^{2}$ assuming the values obtained for the asymmetries are
valid at that scale. We also have evolved the parametrizations using
spin dependent Altarelli-Parisi evolution equations obtaining asymmetries
with a very mild $Q^{2}$ dependence, as it was reported in the experimental
analysis of the data.
A Fortran subroutine that gives the resulting spin dependent parton
distributions
is available upon request\footnote{DEFLO@venus.fisica.unlp.edu.ar}.

\begin{center}
\begin{tabular}{|c|c|c|}\hline
{\small Parameter} &{\small Set LP1} & {\small Set LP2}\\ \hline
$a_{u_{v},0}=a_{\bar{u}}$ &$0.150$&$0.170$ \\
$a_{u_{v},1}$ &$0.013$&$0.003$ \\
$a_{d_{v},0}=a_{\bar{d}}$ &$0.600$&$0.600$ \\
$a_{d_{v},1}$ &$0.100$&$0.100$ \\
$a_{\bar{s}}$&$10a_{\bar{u}}$&$0.050$ \\
$a_{g}$&$0.055$&$-$ \\ \hline
\end{tabular}
\\\vspace*{10mm}{\large \bf Table 1.}\\
\end{center}

Table (2) compares the $\chi^{2}$ values obtained with different sets.
It is clear form the table that it is not possible to discriminate
between our sets using only polarised deep inelastic scattering data.
The parametrizations taken from references \cite{elliot} and \cite{exp}
contain both a large amount of gluon polarisation and were proposed before
the recent data and reanalysis were published, nevertheless they yield
sensible $\chi ^{2}$ values.
It is worth noticing that a half of the total $\chi ^{2}$ in our sets comes
from the comparision with SMC proton data at values of $x$ not very small,
where the sets give good account of the other experiments.

\begin{center}
\begin{tabular}{|c|c|c|c|c|c|}\hline
{\small Experiment} &{\small N$^{0}$ of} & \multicolumn{4}{c|}{\small $\chi
^{2}$}\\ \cline{3-6}
                    &{\small data      } & {\small LP1}& {\small LP2} & {\small
Ref.[15]} & {\small Ref.[16]} \\ \hline
E80 $A_{1}^{p}$     &           4        &   1.43      &   1.53  &   3.56     &
 2.78   \\
E130 $A_{1}^{p}$    &           8        &   3.43      &   3.41  &  13.12     &
 4.45   \\
EMC $A_{1}^{p}$     &          10        &   3.94      &   3.89  &   8.64     &
 9.50   \\
SMC $A_{1}^{d}$     &          11        &   5.65      &   5.60  &   4.63     &
 5.41   \\
E142 $A_{1}^{n}$    &           8        &   3.27      &   3.38  &   5.86     &
 5.42   \\
SMC $A_{1}^{p}$     &          12        &  22.98      &  20.89  &  22.33     &
 32.52  \\ \hline
Total               &          53        &  40.07      &  38.07  &  58.14     &
 60.08  \\ \hline
\end{tabular}
\\\vspace*{10mm}{\large \bf Table 2.}\\
\end{center}
In Table (3) we show values for the moments  $\Gamma_{1}$ obtained
integrating the distributions together with those reported by the
experimental collaborations (assuming an extrapolation for the low $x$
contributions), and the sum rule expectations (corrected because of
the anomaly).

\begin{center}
\begin{tabular}{|c|c|c|c|c|c|c|}
\hline
 &Experimental & Sum rule  &\multicolumn{4}{c|}{Parametrizations} \\
\cline{4-7}
 & Data        & prediction&{\small LP1}&{\small LP2}&{\small Ref.[15]}&{\small
Ref.[16]} \\ \hline
 $\Gamma_{1}^{p}-\Gamma_{1}^{n}$&\footnotesize{SMC$+$EMC} $0.204\pm0.029$ &
0.193& $0.194$ & $0.194$ & $0.205$ & $0.210$ \\ \hline
$\Gamma_{1}^{d}$&\footnotesize{SMC}        $0.023\pm0.020$ &0.035& $0.034$ &
$0.037$ & $0.033$ & $0.030$ \\ \hline
$\Gamma_{1}^{p}$&\footnotesize{SMC}        $0.136\pm0.011$ &0.136& $0.135$ &
$0.137$ & $0.139$ & $0.139$ \\ \hline
$\Gamma_{1}^{n}$&\footnotesize{E142}      $-0.031\pm0.011$ &-0.056& $-0.059$ &
$-0.057$ & $-0.066$ & $-0.071$ \\ \hline
$F/D$                                      & $0.573\pm0.01$ &-& $0.577$ &
$0.576$ & $0.578$ & $0.549$ \\ \hline
$F+D$                                      & $1.257\pm0.003$ &-& $1.265$ &
$1.265$ & $1.357$ & $1.375$ \\ \hline
\end{tabular}
\\\vspace*{10mm}{\large \bf Table 3.}\\
\end{center}
The only line for which there seems to be a disagreement between our sets
and the quoted experimental values is the one for $\Gamma_{1}^{n}$.
As our sets give a good account of E142 data, the discrepancy seems to lay
in the small $x$ behaviours assumed (higher twists are negligible at
$Q^{2}=10\,GeV^{2}$ where the analysis is performed). In the case of E142 data,
the
extrapolation begins at a rather high value of $x$, ($x=0.03$), so the
difference can be substantial. This is in fact the case, as it is shown in
Table (4) where the contributions from the unmeasured regions are tabulated.
\begin{center}
\begin{tabular}{|c|c|c|c|}
\hline
 & Value from &\multicolumn{2}{c|}{From parametrizations} \\ \cline{3-4}
 & extrapolation   &{\small  LP1}&{\small LP2} \\ \hline
 $\Gamma_{1}^{p}-\Gamma_{1}^{n}\mid_{x<0.006}$&\footnotesize{SMC$+$EMC}
$0.007\pm0.007$ & $0.008$ & $0.010$  \\ \hline
$\Gamma_{1}^{d}\mid_{x<0.006}$&\footnotesize{SMC}        $-0.003\pm0.003$ &
$-0.006$ & $-0.003$  \\ \hline
$\Gamma_{1}^{n}\mid_{x<0.03}$&\footnotesize{E142}        $-0.009\pm0.006$ &
$-0.031$ & $-0.030$  \\ \hline
$\Gamma_{1}^{p}\mid_{x<0.003}$&\footnotesize{SMC}        $0.004\pm0.002$ &
$-0.002$ & $0.002$  \\ \hline
\end{tabular}
\\\vspace*{10mm}{\large \bf Table 4.}\\
\end{center}

There is a significative difference between the value coming form the
extrapolation assumed by E142  and those produced by the behaviour of the
distributions in our sets. We remind the reader that this behaviour, in the
spin
dilution model, depends on three factors; the actual behaviour of the
unpolarised
parton distributions, the constraints on the spin dependent distributions,
and the parameters fixed by the available data, which include the lower $x$
data of the other experiments.

Between the two sets there is a slight difference arising form the dominance
of either the gluons or the sea quarks, illustrated by the low $x$ contribution
to the
proton moment. This contribution is negative in the LP1 set (dominated by
gluons)
whereas is positive in the LP2, where a conspiracy between the sea
contributions
yields the result.

Summarizing this section, we conclude that it is possible to bild sensible
spin dependent parton distributions compatible with all polarised deep
inelastic
scattering experiments. Allowing certain degree of gluon or strange
quark polarisation, considerably smaller than originally thought, the
distributions
are also compatible with the Bjorken and Ellis-Jaffe sum rules and
$\beta$-decay
data. The allegued discrepancy between the different experiments, and also
between experimental results and sum rule predictions, is shown to be due to
the inconsistent way in which the data is extrapolated to small $x$.
Deep inelastic scattering experiments are not, however, able to discriminate
between sets with gluon or sea quark polarisation. In the next section
we show that the discrimination can be done measuring spin-spin asymmetries
in polarised proton-proton collisions.

\pagebreak
\noindent {\large \bf III. Spin-spin asymmetries:}\\

Several processes in which, in principle, the sea quark and gluon
polarisation can be extracted directly from experiment have been suggested
since the so called proton spin crisis began \cite{Rey}.
In this section we compare predictions for heavy quark pair production, direct
photon production at large $p_{T}$ and Drell-Yan processes in polarised
proton-proton colisions coming from sets of spin dependent quark and gluon
distributions with different assumptions about the polarisation of sea quarks
and gluons. As these processes are much more sensitive to the sea quark and
gluon disributions than deep inelastic scattering, they will be able to
complement the information obtained up to now.

We begin considering the Drell-Yan proton-proton polarised asymmetry defined
by
\begin{equation}
A_{LL}^{DY}=\frac{d\sigma \uparrow \uparrow/dQ^{2}-d\sigma \uparrow
\downarrow/dQ^{2}}
{d\sigma \uparrow \uparrow/dQ^{2}+d\sigma \uparrow \downarrow/dQ^{2}}
\end{equation}
where $d\sigma \uparrow\uparrow$ ($d\sigma \uparrow\downarrow$) denotes the
cross section for the configuration where the incoming proton spins are
parallel (antiparallel), and $Q^{2}$ is the invariant mass squared of the
outgoing lepton pairs.

It has been suggested \cite{exp} that this asymmetry is particularlly useful
to discern between a large gluon polarisation and a large polarisation of the
sea quarks. The argument given
in reference \cite{exp} for this is that $A_{LL}^{DY}$, calculated with
a set of parton distributions where the sea is negatively polarised, is
positive
whereas, when calculated using a set with large gluon polarisation, is
negative.
It is, then, just a question of measuring the sign of this asymmetry to ascribe
the defect in $\Gamma_{1}^{p}\mid_{EMC-SMC}$ either to $\Delta s$ or $\Delta
G$.

Performing the computations with our sets we find, however, that the Drell-Yan
experiment is unable to discriminate between the two different scenarios.
The difference in signs obtained in reference \cite{exp}
is just a consequence of having mixed different factorization schemes
in the analysis. On one side they use a cross-section defined for a DIS
scheme-like distribution \cite{Rat}, however their distributions are extracted
in the Altarelli-Ross scheme \cite{AA}. Using the relation between both
schemes given in Eq.(12), it is straightforward to write the Drell-Yan cross
section
in the AR scheme
\begin{eqnarray}
{{d\Delta \sigma^{DY}}\over {dQ^2}}&=&-{{4\pi \alpha^2}\over {9sQ^2}}
\int^1_0  {{dx_1}\over {x_1}} \int^1_0
{{dx_2}\over {x_2}}\sum_i \left\{ \left[ e^2_i\Delta q^i(x_1,t)\Delta \bar
q^i(x_2,t) + (1\leftrightarrow 2)
\right] \right. \nonumber \\
&&\left. \times \left[\delta (1-z)  + \theta (1-z)
\alpha_s \Delta w_q (z)  \right]
 + (e_i^2 \left[ \Delta q^i(x_1,t) +\Delta \bar q^i(x_1,t) \right]\right.
\nonumber \\
&& \left. \times \Delta G(x_2,t)
 + (1 \leftrightarrow 2 )
)\left[-{{\alpha_s}\over {4\pi}} \delta (1-z)+  \theta (1-z) \alpha_s \Delta
w_G (z)\right]  \right\}
\end{eqnarray}
where
\begin{eqnarray}
\Delta w_q (z)&=& {4\over {6\pi}}\left[ (1+{4
\over 3}\pi^2)\delta (1-z)+ {3\over{(1-z)_+}}+2(1+z^2)\left[{{\log
(1-z)}\over{1-z}} \right]_+ -4-2z\right] \nonumber \\
\Delta w_G (z)&=& {1\over {4\pi}}\left[ (2z-1)\log
(1-z) -{3\over 2}z^2+3z-{1\over 2}\right]
\end{eqnarray}
and
\begin{equation}
z=\frac{\tau}{x_{1}x_{2}}=\frac{Q^{2}}{sx_{1}x_{2}}
\end{equation}
This cross section has an extra ``delta function" term proportional to the
polarised gluon distribution which makes the quark and gluon next to leading
order term to be equally important. In Figures (4) and (5) we compare the
$A_{LL}^{DY}$
calculated as in reference \cite{exp} with the corrected predictions.
Both Figures correspond to $\sqrt{s}=27\,GeV$. The
correction changes the sign of the asymmetry in the gluonic scenario and is
relatively small for the other at values of $Q^{2}$ not very high.
The correction clearly reduces the diferences between the predictions coming
from
the two scenarios.

There exist also an ambiguity related to the way the spin
is distributed among the flavours of the sea which reduces also the
discriminative
power of the experiment the gluon-strange quark alternatives but allows another
application. In the sets proposed in reference \cite{exp} the sea
polarisation is $SU(3)$ invariant ($\Delta u=\Delta d= \Delta s$). Another
posibility, implemented in our sets,
is that the non strange sea is polarised parallel to the net valence
polarisation
and only the the strange quarks became negatively polarised (in set 2).
Predictions
for $A_{LL}^{DY}$ using our sets are shown in Figure (6).  There is a
negligible
difference between the predictions of both sets due not only to the fact
that in the new sets the gluons or the strange quarks are less polarised but
to the fact that the Drell-Yan asymmetry picks also (and prevailingly)
the non strange sea. We therefore conclude that this asymmetry
can not tell us whether gluons contribute to $g_{1}^{p}$ or not but
can discriminate between an $SU(3)$ symmetric sea ($A_{LL}^{DY}>0$)
and one where this symmetry is broken ($A_{LL}^{DY}<0$).

Another candidate for probing the sea and gluon polarisation is
the direct photon production in proton-proton collisions \cite{Ber}. At
parton level, and to the lowest order, prompt photons are produced via
Compton scattering $qG\rightarrow \gamma q$ and annihilation $q\overline{q}
\rightarrow \gamma G$.
In the leading order, the differential cross section reads
\begin{eqnarray}
\lefteqn{E_{\gamma}d^{3}\Delta \sigma/dp^{3}_{\gamma} =} \nonumber \\
& & \frac{\alpha_{em} \alpha_{s}}{s}
\int_{x_{min}}^{1}dx_{1}\frac{1}{x_{1}x_{2}(x_{1}s+u)}[\sum_{i}^{n_{f}}e_{i}^{2}
[\Delta q_{i}(x_{1},Q^{2}) \Delta \overline{q}_{i}(x_{2},Q^{2})+(1
\leftrightarrow 2)]
\frac{d \Delta \hat{\sigma}}{d\hat{t}}(q\overline{q} \rightarrow \gamma G)
\nonumber \\
& & +2g_{1}(x_{1},Q^{2})\Delta G(x_{2},Q^{2})
\frac{d \Delta \hat{\sigma}}{d\hat{t}}(q G \rightarrow \gamma q)
+ 2\Delta G(x_{1},Q^{2})g_{1}(x_{2},Q^{2})\frac{d \Delta
\hat{\sigma}}{d\hat{t}}(q G \rightarrow \gamma q)]
\end{eqnarray}
where
\begin{eqnarray}
\frac{d \Delta \hat{\sigma}}{d\hat{t}}(q \overline{q} \rightarrow \gamma G) & =
&
- \frac{8}{9}[\frac{\hat{t}}{\hat{u}}+\frac{\hat{u}}{\hat{t}}] \nonumber \\
\frac{d \Delta \hat{\sigma}}{d\hat{t}}(q G \rightarrow \gamma q) & = &
- \frac{1}{3}[-\frac{\hat{t}}{\hat{s}}+\frac{\hat{s}}{\hat{t}}] \\
\frac{d \Delta \hat{\sigma}}{d\hat{t}}(G q \rightarrow \gamma q) & = &
- \frac{1}{3}[-\frac{\hat{u}}{\hat{s}}+\frac{\hat{s}}{\hat{u}}] \nonumber
\end{eqnarray}
and
\begin{equation}
\hat{s}=x_{1}x_{2}s, \,\,\, \hat{t}=x_{1}t, \,\,\, \hat{u}=x_{2}u
\end{equation}
The cross section has been calculated up to two loops in references
\cite{loop1} \cite{loop2}in different factorization schemes. There, it has
been shown that the resulting corrections for the corresponding
asymmetry are small.

The asymmetries calculated with  our two sets
at $\sqrt{s}=100\,GeV$ and $p_{T}=5\, GeV$ are shown in Figure (7).
In this case we find a clean difference between the prediction of both sets.
As the Compton subprocess dominates
over the annihilation one, this experiment avoids, up to a certain extent, the
ambiguity related to the way in which the sea is polarised.

Heavy quark production in polarised proton-proton collisions is another
experiment dominated by gluon-gluon fusion and, consequently,
can corroborate the extraction of $\Delta G$ coming from the previous one.
Following reference \cite{j/psi}, we calculate the $J/\Psi$ production
two spin asymmetry
\begin{equation}
A_{LL}^{J/\Psi}=\frac{d\sigma \uparrow \uparrow/d^{3}p-d\sigma \uparrow
\downarrow/d^{3}p}
{d\sigma \uparrow \uparrow/d^{3}p-d\sigma \uparrow \downarrow/d^{3}p}
=\frac{E d\Delta\sigma /d^{3}p}{Ed\sigma /d^{3}p}
\end{equation}
where
\begin{eqnarray}
E\frac{ d\Delta\sigma}{d^{3}p} & = & \frac{1}{\pi} \int_{x_{1}^{min}}
^{1}dx_{1} \Delta G(x_{1},Q^{2}) \Delta G(x_{2},Q^{2})(\frac{x_{1}x_{2}}{x_{1}-
\frac{e^{y}}{\sqrt{s}}\sqrt{m^{2}_{J/\Psi}+p_{T}^{2}}})\frac{d\Delta
\hat{\sigma}}{d\hat{t}}\\
E\frac{ d\sigma }{d^{3}p} & = & \frac{1}{\pi} \int_{x_{1}^{min}}
^{1}dx_{1}  G(x_{1},Q^{2})  G(x_{2},Q^{2})(\frac{x_{1}x_{2}}{x_{1}-
\frac{e^{y}}{\sqrt{s}}\sqrt{m^{2}_{J/\Psi}+p_{T}^{2}}})\frac{d
\hat{\sigma}}{d\hat{t}}\\
\end{eqnarray}
and $d \hat{\sigma}/d\hat{t}$  is the differential cross section for the
subprocess $GG \rightarrow J/\Psi \,G$,
using sets LP1 and LP2. Figure (8) shows the prediction computed at
$\sqrt{s}=20\, GeV$,
 $y=0$, as a function of $p_{T}$.
For the kinematical range $1<p_{T}<6 \,GeV$ the predictions differ
substantially and can
be eventually discriminated. For higher values of $p_{T}$ both sets yield
similar results
being the cross section dominated by gluons coming from the $Q^{2}$ evolution.

We conclude this section pointing up how polarised proton-proton collisions
would be able to test the assumptions made in the previous section providing
valuable pieces of information about polarised parton distribution.

The $J/\psi$ production asymmetry is proportional to the gluon spin-dilution
function squared so it not only measures the net gluon polarisation $\Delta G$
but the explicit $x$ dependence, which is assumed in the fits.
The Drell Yan asymmetry tells us whether the SU(3) symmetry in the sea is a
good approximation or not, yielding constraints on the $x$ dependence and
normalization  of the polarised sea distributions. Finally, direct photon
production, being proportional to the products of valence quark and gluon
spin-dilution functions and valence quarks and sea quarks in different
kinematical regions, allows a cross check for the information obtained
from all the above mentioned polarised experiments. \\

\noindent {\large \bf IV. Conclusions:}\\

Recent data on proton polarised asymmetry in deep inelastic scattering
reported by the SMC collaboration indicate a substantial reduction
in the amount of gluon or strange quark polarisation needed to bring agreement
between the Ellis-Jaffe prediction for $\Gamma_{1}^{p}$ and its experimental
value. This reduction has important consequences in the obtention of
spin dependent parton distributions, which inevitably includes assumptions
on the size of these quantities, and the feasibility of some experiments
intended to measure them.

In this paper we have constructed two different sets of parton distributions
in the framework of the spin dilution model with the amount of gluon or strange
quark polarisation suggested by the most recent experiments.
We have analysed the consequences of this reduction in the global fit
of spin dependent data finding a mild improvement but not significative.
We also have examined predictions for three different polarised proton-proton
collission
cross sections concluding that whereas the Drell-Yan spin-spin asymmetry
is not a good test to see whether the gluons or the strange quarks contribute
dominantly, prompt photon and $J/\Psi$ production  in proton-proton
collisions can measure the size of the gluon polarisation.

\pagebreak

\pagebreak

\noindent{\large \bf Figure Captions}
\\

\begin{enumerate}
\item[Figure 1 ]  The spin-dependent proton asymmetry given by the model
(set LP1)  compared to SMC \cite{SMC2}, EMC \cite{EMC} and earlier SLAC
data \cite{data}.
\item[Figure 2 ] The same as Figure (1) but for the spin-dependent neutron
asymmetry given by E-142 \cite{E142}.
\item[Figure 3 ] The same as Figure (1) but for the spin-dependent deuteron
asymmetry given by SMC \cite{SMC}.
\item[Figure 4 ] The Drell-Yan spin-spin asymmetry as calculated in reference
\cite{exp} and corrected.
\item[Figure 5 ] The Drell-Yan spin-spin asymmetry calculated with sets LP1
and LP2.
\item[Figure 6 ] The direct photon spin asymmetry calculated with sets LP1
and LP2.
\item[Figure 7 ] The $J/\Psi$ production two spin asymmetry calculated with
sets LP1 and LP2.

\end{enumerate}

\end{document}